\documentclass[amsmath,amssymb,showpacs,draft,preprint,floatfix]{revtex4}
\usepackage{graphicx}
\usepackage{latexsym}
\usepackage{amsmath}
\usepackage{amsfonts}
\usepackage{amssymb}
\begin{document}

\title{Mirror-field-atom interaction: Hamiltonian diagonalization}

\author{D. Rodr\'{\i}guez-M\'endez,$^1$ O. Aguilar-Loreto,$^2$
H. Moya-Cessa,$^1$  }
\affiliation{$^1$INAOE, Apdo. Postal 51 y 216, 72000, Puebla, Pue., Mexico \\
$^2$ Departamento de Ingenierias, CUCSur, Universidad de
Guadalajara C.P. 48900, Autl\'an de Navarro, Jal., Mexico  }

\begin{abstract}
We show that the interaction between a movable mirror with a
quantized field that interacts with a two-level atom may be
simplified via a transformation that involves Susskind-Glogower
operators (SGO). By using this transformation it is easy to show
that we can cast the Hamiltonian, after a series of small
rotations, into an effective Hamiltonian  that may be solved. We
would like to stress that the transformation in terms of SGO
already simplifies enough the Hamiltonian in the sense that, in an
exact way, it "eliminates" one of the three-subsystems, namely the
quantized field.
\end{abstract}
  \maketitle
\section{Introduction}
Recently, special attention has been devoted to a system
consisting of a cavity field and a movable
mirror\cite{Meystre,Li}. This is due to the fact that for such a
system we can produce non-classical states \cite{Peter},
particularly the macroscopic superposition of at least two
coherent states, i.e. Schr\"odinger cat-states. The concept of
superposition of states plays a fundamental role in understanding
the foundations of quantum mechanics, this is why the generation
of  non-classical states, such as squeezed states \cite{Guasti},
and the particularly important limit of extreme squeezing, i.e.
Fock or number states \cite{Krause}, has been widely studied in
several systems. It is known that a non-linear interaction can
generate Schr\"odinger cat-states. The non-linear interaction used
to generate such states is the one produced by a Kerr medium
\cite{Agarwal,Tanas} which corresponds to a quadratic Hamiltonian
in the number field operator \cite{manci,Lara}.  Our main
motivation to make the field-mirror system interact with an atom
is to look for the possibility to extract information about it by
later measuring atomic properties. This, because it is well known
that several quasiprobability reconstruction techniques
\cite{Roversi1,Roversi2} for the quantized field \cite{lut} or the
vibrational motion of an ion \cite{Tomb1,Tomb2}, rely on the
measurement of atomic states. Therefore, the passage of atoms
through such systems, could give us information, not only about
the states of the mirror or field, but about its dynamical
interaction. This is, the passage of a two-level atom through a
cavity containing a movable mirror may give us information about
the entanglement between  mirror and field. The purpose of this
contribution is not to study this possibility, however, but to
show how the total system may be simplified, by several rotations
(some of them small rotations, i.e. approximations) that
diagonalize the Hamiltonian in such a way that a solution may be
easily obtained. We will study the possibility of reconstructing
the mirror-field interaction elsewhere.

\section{Interaction between the cavity and the mirror}
 The
interaction between an electromagnetic field and a movable mirror
(treated quantum mechanically)  has a relevant Hamiltonian given
by \cite{manci} (we set $\hbar=1$)

\begin{equation}
H_{f-m}=\omega a^{\dagger}a + \nu b^{\dagger}b -g
a^{\dagger}a(b^{\dagger}+b) ,\label{ham}
\end{equation}
where $a$ and $a^{\dagger}$ are the annihilation and creation
operators for the cavity field, respectively. The field  frequency
is $\omega$. $b$ and $b^{\dagger}$ are the annihilation and
creation operators for the mirror oscillating at a frequency $\nu$
and
\begin{equation}
g = \frac{\omega}{L}\sqrt{\frac{\hbar}{2m\nu}},
\end{equation}
with $L$ and $m$ the length of the cavity and the mass of the
movable mirror.
\section{Mirror-Field-Atom interaction}
If we pass a two-level atom through a cavity with a movable mirror
as the one described by equation (\ref{ham}), we have have to add
the free Hamiltonian for the atom and the interaction with the
quantized field, so we obtain \cite{chin}
\begin{eqnarray}
H_{a-f-m}&=& \frac{\omega_0}{2} \sigma_z+\lambda\left(
a\sigma_{+}+a^{\dag}\sigma_{-}\right)+\omega a^{\dagger}a + \nu
b^{\dagger}b \nonumber \\
& -&g a^{\dagger}a(b^{\dagger}+b), \label{ham2}
\end{eqnarray}
where $\lambda$ is the atom-field interaction constant, $\omega_0$
is the atomic transition frequency and $\sigma_{-}$ ($\sigma_{+}$)
is the lowering (raising) operator for the atom, with
$[\sigma_{+}, \sigma_{-}]=2\sigma_{z}$.

We consider the on-resonant interaction between the field an the
atom, i.e. $\omega=\omega_0$, and pass to an interaction picture,
taking advantage that the operator $\omega(
a^{\dagger}a+2\sigma_z/2)$ commutes with all the other operators
involved in the Hamiltonian, to obtain
\begin{equation}
\widehat{H}=\nu\widehat{N}+\chi\widehat{n}\left( b+b^{\dag}\right)
+\lambda\left( a\sigma_{+}+a^{\dag}\sigma_{-}\right).
\label{total}
\end{equation}
The quantities $\widehat{n} =a^{\dag}a$ and $\widehat{N}
=b^{\dag}b$ are the number operators for the field and mirror,
respectively. We will use the Susskind-Glogower operators
\cite{Suskind}
\begin{equation}
V= \frac{1}{\sqrt{\hat{n}+1}}a, \qquad
V^{\dag}=a^{\dagger}\frac{1}{\sqrt{\hat{n}+1}},
\end{equation}
that satisfy the commutation relation $[V,V^{\dag}]=|0\rangle
\langle 0|$ to  transform the above Hamiltonian with the following
matrix operator \cite{lasphys}
\begin{equation}
M^{\dag}=\left(
\begin{array}
[c]{cc}%
1 & 0\\
0 & V^{\dag}%
\end{array}
\right)  ,\qquad M=\left(
\begin{array}
[c]{cc}%
1 & 0\\
0 & V
\end{array}
\right),\label{maint} \end{equation} such that we can rewrite the
interaction Hamiltonian as
\begin{equation}
\widehat{H}=\widehat{H}_{V}+\widehat{\rho}_{22}^{0}%
\end{equation} where
\begin{eqnarray}&&\widehat{H}_{V}=M^{\dag}
 \\ &&
  \left(
\begin{array}
[c]{cc}%
\nu\widehat{N}+\chi\widehat{n}\left(  b+b^{\dag}\right)  & \lambda
\sqrt{\widehat{n}+1}\\
\lambda\sqrt{\widehat{n}+1} & \nu\widehat{N}+\chi\left(
\widehat{n}+1\right) \left(  b+b^{\dag}\right)
\end{array}
\right)  M ,\nonumber
\end{eqnarray}
and
\[
\widehat{\rho}_{22}^{0}=\left(
\begin{array}
[c]{cc}%
0 & 0\\
0 & \nu\widehat{N}%
\end{array}
\right) . \left\vert 0\right\rangle \left\langle 0\right\vert
\]
Note that%
\[
\left[  \widehat{H}_{V},\rho_{22}^{0}\right]  =0,
\]
therefore we can write the evolution operator as
\begin{equation}
\widehat{U}\left(  t\right)  =e^{-i\widehat{H}t}=e^{-i\widehat{H}_{V}%
t}e^{-i\widehat{\rho}_{22}^{0}t}.%
\end{equation}
In order to calculate $e^{-i\widehat{H}_{V} t}$ , we develop the
exponential in Taylor series take into account that
\begin{eqnarray}&& \widehat{H}_{V}^k=M^{\dag} \nonumber \\
&&  \left(
\begin{array}
[c]{cc}%
\nu\widehat{N}+\chi\widehat{n}\left(  b+b^{\dag}\right)  & \lambda
\sqrt{\widehat{n}+1}  \nonumber\\
\lambda\sqrt{\widehat{n}+1} & \nu\widehat{N}+\chi\left(
\widehat{n}+1\right) \left(  b+b^{\dag}\right)
\end{array}
\right)^k  M, \\&& k\ge 1,
\end{eqnarray}
so we write the evolution operator as
\[
\widehat{U}\left(  t\right)  =M^{\dag}e^{-i\widehat{\widetilde{H}}%
t}Me^{-i\widehat{\rho}_{22}^{0}t}+\left(
\frac{1-\sigma_{Z}}{2}\right)
\left\vert 0\right\rangle \left\langle 0\right\vert e^{-i\widehat{\rho}%
_{22}^{0}t},%
\]
where
\begin{equation}
\widehat{\widetilde{H}}=\left(
\begin{array}
[c]{cc}%
\nu\widehat{N}+\chi\widehat{n}\left(  b+b^{\dag}\right)  & \lambda
\sqrt{\widehat{n}+1}\\
\lambda\sqrt{\widehat{n}+1} & \nu\widehat{N}+\chi\left(
\widehat{n}+1\right) \left(  b+b^{\dag}\right)
\end{array}
\right) .%
\end{equation}%
Up to here we have realized already a relevant simplification:
this Hamiltonian, unlike the one in equation (\ref{total}), has
field operators in it that commute with each other, and, because
they commute with the other sub-systems operators, they may be
treated from now on as classical numbers. Therefore, we have
effectively and exactly eliminated one sub-system, namely the
field, from the initial problem.

Now we will take advantage of the difference in order of
magnitudes (the atom-field interaction constant is much larger
than the mirror-field interaction constant) of the different
constants in this interaction to produce an effective Hamiltonian
which being diagonal, may be solved exactly. For this we will use
a small rotation approach proposed by Klimov and S\'anchez-Soto
\cite{kli}. First, let us do an exact rotation to the Hamiltonian
and obtain
\[
R=\frac{1}{\sqrt{2}}\left(
\begin{array}
[c]{cc}%
1 & 1\\
-1 & 1
\end{array}
\right),
\]
with%
\[
H_{R}=R\widetilde{H}R^{\dag},%
\]
so that we have
\begin{eqnarray}
\nonumber \widehat{H}_{R}&=&\nu\widehat{N}+\chi\left(
\widehat{n}+\frac{1}{2}\right) \left(  b+b^{\dag}\right)
+\lambda\sigma_{z}\sqrt{\widehat{n}+1}\\&+&\frac{\chi }{2}\left(
\sigma_{+}+\sigma_{-}\right)  \left(  b+b^{\dag}\right).
\label{rotado2}%
\end{eqnarray}
Now we apply the following small rotations
\begin{align*}
\widehat{U}_{1}  &  =e^{\xi_{1}\left(
b^{\dag}\sigma_{+}-b\sigma_{-}\right)
},\\
\widehat{U}_{2}  &  =e^{\xi_{2}\left(
b\sigma_{+}-b^{\dag}\sigma_{-}\right)
},%
\end{align*}
we call the small because we will consider (see below)
$\xi_{1},\xi_{2}\ll 1$, i.e. they will perform tiny rotations.
Transforming (\ref{rotado2}) with the above operators we obtain
\[
H_{2}=U_{2}U_{1}H_{R}U_{1}^{\dag}U_{2}^{\dag}.%
\]
Due to the fact that the $\xi$ 's are small we remain to first
order when we develop the exponentials in Taylor series so that we
finally obtain
\begin{align*}
\widehat{H}_{2}  &  \approx \nu\widehat{N}+\lambda\sqrt{\widehat{n}+1}\sigma_{z}%
+\chi\left(  \widehat{n}+\frac{1}{2}\right)  \left(  b+b^{\dag}\right) \\
&  +\left(  \xi_{1}+\xi_{2}\right)  \chi\sigma_{z}\left(
b+b^{\dag}\right)
^{2}\\
&  +\left(  \xi_{2}-\xi_{1}\right)  \left(  \frac{\chi}{2}\left(
\sigma _{+}+\sigma_{-}\right)  ^{2}+\chi\left(
\widehat{n}+\frac{1}{2}\right)
\left(  \sigma_{+}+\sigma_{-}\right)  \right) \\
&  +\left(  \frac{\chi}{2}-\xi_{1}\left(  \nu+\lambda\sqrt{\widehat{n}%
+1}\right)  \right)  \left(  b^{\dag}\sigma_{+}+b\sigma_{-}\right) \\
&  +\left(  \frac{\chi}{2}+\xi_{2}\left(  \nu-\lambda\sqrt{\widehat{n}%
+1}\right)  \right)  \left(
b\sigma_{+}+b^{\dag}\sigma_{-}\right).
\end{align*}
We can choose
\begin{align*}
\xi_{1}\left(  \widehat{n}\right)   &  =\frac{\chi}{2\left(
\nu+\lambda
\sqrt{\widehat{n}+1}\right)  },\\
\xi_{2}\left(  \widehat{n}\right)   &  =-\frac{\chi}{2\left(
\nu-\lambda
\sqrt{\widehat{n}+1}\right)  },%
\end{align*}
so that the last two terms of  $\widehat{H}_{2}$ become zero. For
the other terms we need to calculate
\begin{equation}
\xi_{1}+\xi_{2}
=\frac{\lambda\chi\sqrt{\widehat{n}+1}}{\lambda^{2}\left(
\widehat{n}+1\right)  -\nu^{2}},
\end{equation}
and using the fact that the $\xi$'s are small we can also neglect
the term
\begin{equation}
\xi_{2}-\xi_{1}    =\frac{\chi\nu}{\left(  \lambda^{2}\left(
\widehat {n}+1\right)  -\nu^{2}\right)
}\sim\frac{1}{\lambda^{2}}\sim 0.
\end{equation}
Taking into account that
\begin{equation}
\frac{\chi^{2}}{\lambda^{2}\left( \widehat{n}+1\right)  -\nu^{2}}
\approx\frac{\chi^{2}}{\lambda^{2}\left(  \widehat{n}+1\right)  }%
\end{equation}
we may finally write
\begin{eqnarray}
\nonumber \widehat{H}_{2}&&\approx\nu\widehat{N}+\chi\left(  \widehat{n}+\frac{1}%
{2}\right)  \left(  b+b^{\dag}\right)
+\lambda\sqrt{\widehat{n}+1}\sigma _{z}
\\&&+\frac{\chi^{2}}{\lambda\sqrt{\widehat{n}+1}}\left(
b+b^{\dag}\right)
^{2} \label{rotado3}.%
\end{eqnarray}
This is a Hamiltonian that is already diagonal and direct to
solve. The purpose of this contribution was to show that the
Hamiltonian of the total interaction could be simplified. We have
achieved this. A more complete study to look for the evolution of
observables and possibilities of reconstructing the mirror-field
interaction is still in preparation and will be published
elsewhere.
\section{Conclusions}
We have shown that the problem of a quantized field interacting
simultaneously with a two-level atom and a movable mirror may be
diagonalized via a set of transformations, the main one, being a
transformation that involves Susskind-Glogower operators, equation
(\ref{maint}). This transformation, besides the fact that does not
involve approximations, allows us to simplify the problem by
"eliminating" the field operators to leave an effective
interaction between atom and mirror. The Hamiltonian for this
interaction then may be slightly rotated to obtain a dispersive
Hamiltonian that being diagonal is already solvable such that the
evolution operator may be eventually found. Here, we treated the
ideal case where the environment was not taken into account. This
may allow us to study elsewhere possibilities of vibrational
purification via multiple interactions of atoms traversing the
cavity \cite{Arancibia}.


\end{document}